\documentstyle[preprint,prb,aps]{revtex}

\tighten

\begin{document}
\draft
\title{Dynamical Properties of the Landau-Ginzburg Model with Long-Range Correlated
Quenched Impurities}
\author{Elka Korutcheva\thanks{%
Permanent address: G.Nadjakov Institute of Solid State Physics, Bulgarian
Academy of Sciences, 72 Tzarigradsko Shaussee Blvd, 1784 Sofia, Bulgaria}
and F. Javier de la Rubia}
\address{Dept. de F\'{\i}sica Fundamental, Universidad Nacional de \\
Educaci\'{o}n a Distancia, Apdo. de Correos 60.141, 28080 Madrid, Spain}
\date{\today}
\maketitle

\begin{abstract}
We investigate the critical dynamics of the time-dependent Landau-Ginzburg
model with non conserved n-component order parameter (Model A) in the
presence of long-range correlated quenched impurities. We use a special kind
of long-range correlations, $g(|r - r^{\prime}|)\sim |r-r^{\prime}|^{-a}$, $%
a = 4-\delta$, previously introduced by Weinrib and Halperin. The parameter $%
\delta \sim \epsilon$, where $\epsilon = 4 - d$ is the usual small parameter
within the Wilson-Fisher renormalization group theory.

Using a double expansion in $\epsilon $ and $\delta $ we calculate the
critical exponent $z$ up to second order on the small parameters. We show
that the quenched impurities of this kind affect the critical dynamics
already in first order of $\epsilon $ and $\delta $, leading to a relevant
correction for the mean field value of the exponent $z$.
\end{abstract}


The static and dynamic scaling properties of isotropic systems with
short-range (SR) correlated quenched impurities has been the subject of a
large number of studies, and is now well understood \cite
{Lub,Khm,GL,GMM,UKM,ED,Ma}. From those works, it is known that the critical
behavior, described by the SR fixed point (f.p.) of the renormalization
group (RG) recursion relations, is consistent with the Harris criterion \cite
{Harris}, which states under which conditions a random critical-temperature
system undergoes a second order phase transition with the same critical
properties as the corresponding pure system.

For the static properties, the case of long-range (LR) correlations was
considered by Weinrib and Halperin \cite{WH}. They introduced a special kind
of spatial Gaussian correlations between the quenched impurities falling off
with the distance as a power law. Their model permits to find a new f.p. of
the RG transformation corresponding to a second order phase transition with
specific critical exponents due to the presence of this kind of impurities.

When disorder with LR power-low correlations is taken into account, the
Harris criterion is modified. If $a$ is the exponent of the LR correlations
(see below), $d$ is the dimensionality of the system, and $\nu $ is the
correlation-length exponent $\xi \sim |T-T_{c}|^{-\nu }$, the modified
Harris criterion \cite{WH} states that the LR disorder is irrelevant if $%
a\nu -2>0$ for $a<d$, while $d\nu -2=-\alpha >0$ for $a>d$ (the normal
Harris criterion).

As a further step in that direction, in this work we analyze, in the
framework of the $\epsilon $-expansion, the critical dynamics of the
time-dependent Landau-Ginzburg (TDLG) model for systems with $n$-component ,
non-conserved order parameter (Model A) \cite{HH}, in the case when the LR
correlations between impurities are of the form introduced by Weinrib and
Halperin. We calculate the dynamic critical exponent $z$ up to second order
in $\epsilon =d_{u}-d$ ($d$ is the dimensionality of the space and $d_{u}=4$
is the upper borderline dimensionality for this model), generalizing the
result of Grinstein, Ma and Mazenko \cite{GMM}, valid for the case of SR
correlation between impurities.

The model we consider is 
\begin{equation}
\frac{\partial {\psi _{\alpha }(\vec{x},t)}}{\partial {t}}=-\Gamma \frac{%
\partial {F}}{\partial {\psi _{\alpha }(\vec{x},t)}}+\eta _{\alpha }(\vec{x}%
,t),  \label{eq1}
\end{equation}
where $\psi _{\alpha }(\vec{x},t),\alpha =1,...,n$ is a $n$-component order
parameter, $\Gamma $ is a bare kinetic coefficient, $\eta _{\alpha }$ is a
Gaussian white noise 
\begin{equation}
<\eta _{\alpha }(\vec{x},t)>_{av}=0,  \label{eq2}
\end{equation}
\begin{equation}
<\eta _{\alpha }(\vec{x},t)\eta _{\beta }(\vec{x^{\prime }},t^{\prime
})>_{av}=2\Gamma \delta (\vec{x}-\vec{x^{\prime }})\delta (t-t^{\prime
})\delta _{\alpha ,\beta }  \label{eq3}
\end{equation}
and 
\begin{equation}
F=\frac{1}{2}\int d^{d}\vec{x}\left[ r\psi ^{2}(\vec{x},t)+(\nabla \psi (%
\vec{x},t))^{2}+\frac{1}{4}u(\psi ^{2}(\vec{x},t))^{2}+\varphi (\vec{x})\psi
^{2}(\vec{x},t)\right]  \label{eq4}
\end{equation}
is the Landau-Ginzburg free energy functional including a static random term 
$\varphi (\vec{x})$, which describes the quenched impurities.

This term obeys a Gaussian distribution

\begin{equation}
<\varphi (\vec{x})>_{av}=0,  \label{eq5}
\end{equation}
\begin{equation}
<\varphi (\vec{x})\varphi (\vec{x^{\prime }})>_{av}=g(|x-x^{\prime }|),
\label{eq6}
\end{equation}
where the function $g(r)$, $r=|\vec{x}-\vec{x^{\prime }}|$ , is a linear
superposition of inverse power law correlations decaying with the distance $%
r $ at different rates $g(r)\sim \sum_{i=1}^{N}A_{i}r^{-a_{i}}$. It is known 
\cite{WH} that of the different components of $g(r)$, the most relevant for
the scaling analysis is that with the slowest decay rate.

The Fourier transform of $g(r)$ for small $k$ is \cite{WH} 
\begin{equation}
g(k)=v+wk^{a-d}.  \label{eq7}
\end{equation}

The introduction of a second small parameter $\delta =4-a$, in addition to $%
\epsilon $, $\delta =O(\epsilon )$, and the double expansion in $\epsilon $
and $\delta $ permits to study the nontrivial LR critical behavior described
by a new f.p. of the RG recursion relations \cite{WH}. This important result
is due to the similar scaling behavior when $\delta \sim O(\epsilon )$ of
the two presumably different (short- and long-range) behaviors implied in (%
\ref{eq7}). Actually, the inequality $a>d$ leads in the long-wavelength
limit $(k\rightarrow 0)$ to the scaling irrelevance of the parameter $w$ and
of any effect of the LR correlations. When $a<d$, the second term in (\ref
{eq7}) is dominant for small $k$, but leading to an unstable critical
behavior \cite{Lub}. Both terms have the same scaling relevance for the
critical behavior only if $a\sim d$ , which will be the case considered here 
\cite{WH}.

The calculation of the critical exponent $z$ closely follows ref.\cite{GMM}.
Using the TDLG equations, (\ref{eq1}--\ref{eq4}) we calculate the
renormalized response function from the Dyson equation, looking for
self-energy $\omega $-dependent contributions of the form $(-i\omega )\ln
(-i\omega )$ \cite{Ma,HH}. A trivial calculation leads to the following
expression for the dynamic exponent $z$ with a correction already in first
order in $\epsilon $ and $\delta $%
\begin{equation}
z=2+(v^{*}+w^{*}),  \label{eq10}
\end{equation}
where $v^{*}$ and $w^{*}$ are the f.p. values corresponding to the LR f.p.
for the parameters $v$ and $w$, derived from the recursion relations to
order $O(\epsilon )$ \cite{WH} 
\begin{equation}
v^{*}=\frac{1}{2(\delta -\epsilon )(5n+4)^{2}}\left( 2(n+2)\epsilon
-(n+8)\delta \right) ^{2},  \label{eq11}
\end{equation}
\begin{equation}
w^{*}=\frac{1}{2(\delta -\epsilon )(5n+4)^{2}}\left( (n+8)\delta
-(2n+4)\epsilon \right) \left( 4(n-1)\delta -3n\epsilon \right) .
\label{eq12}
\end{equation}

For completeness we give here the f.p. value of $u$, which will be used
later 
\begin{equation}
u^{*}=2\frac{(3\delta -\epsilon )}{(5n+4)}.  \label{eq13}
\end{equation}

After substituting the f.p. values $v^{*}$ and $w^{*}$ in Eq.(\ref{eq10}) we
get 
\begin{equation}
z=2+\frac{1}{2(5n+4)}\left( (n+8)\delta -2(n+2)\epsilon \right) ,
\label{eq14}
\end{equation}
valid up to order $O(\epsilon ,\delta )$.

Notice that an $\epsilon $-correction for $z$ is also present in the SR case 
\cite{GMM,Ma}.

The calculation of the second-order correction in $\epsilon $ for the
dynamic exponent $z$ is analogous to that in \cite{GMM}. We again perform
the analysis as before by calculating the $(-i\omega )\ln (-i\omega )$ and $%
(-i\omega )\ln ^{2}(-i\omega )$ contributions up to order $\epsilon ^{2}$.
To this order the response function can be written as

\begin{eqnarray}
G^{-1}(\omega ) &=&\left( \frac{-i\omega }{\Gamma }\right) \left\{ 1+\ln
\left( \frac{-i\omega }{\Gamma }\right) \left[ -\frac{(v+w)}{2}-\frac{3}{8}%
(n+2)u^{2}\ln {\frac{4}{3}}-\frac{1}{2}(v+w)^{2}+\frac{n+2}{4}u(v+w)\right]
\right.   \nonumber \\
&&\left. +\ln ^{2}\left( \frac{-i\omega }{\Gamma }\right) \left[ \left( 
\frac{v+w}{8}\right) \epsilon +\frac{5}{8}(v+w)^{2}-\frac{n+2}{8}u(v+w)-%
\frac{1}{8}w(\epsilon -\delta )\right] \right\} .  \label{response}
\end{eqnarray}

Note that the contribution $-w(\epsilon -\delta )/8$ coming from the diagram
proportional to $v+w$ ensures that the relation
\begin{equation}
\frac{1}{2}\left( \frac{v+w}{2}\right) ^{2}=\left( \frac{v+w}{8}\right)
\epsilon +\frac{5}{8}(v+w)^{2}-\frac{n+2}{8}u(v+w)-\frac{1}{8}w(\epsilon
-\delta )
\end{equation}
is consistently satisfied by the $O(\epsilon )$ f.p. values of the
parameters $u,v$ and $w$ (Eqs.\ref{eq11}-\ref{eq13}). From the response
function we obtain the expression for $z$ up to order $O(\epsilon
^{2},\epsilon \delta ,\delta ^{2})$

\begin{equation}
z=2-\eta +(v+w)+\frac{3}{4}(n+2)u^{2}\ln \frac{4}{3}+\frac{3}{2}(v+w)^{2}-%
\frac{(n+2)}{2}u(v+w),  \label{eq16}
\end{equation}
where the anomalous dimension $\eta $ is \cite{EKDU} 
\begin{equation}
\eta =\frac{1}{16(5n+4)^{2}}\left[ 8n(n+2)\epsilon ^{2}+(n+4)^{2}\delta
(2\epsilon -3\delta )\right] .  \label{eq17}
\end{equation}

From (\ref{eq16}) it is clear that in order to calculate the dynamic
exponent $z$ to order $\epsilon ^{2}$, one needs the expression up to this
order only for the f.p. value of the term $v+w$. For the last three terms,
which are quadratic, it is enough to use the f.p. values of the parameters
given by (\ref{eq11}-\ref{eq13}).

We calculate the f.p. values up to order $\epsilon ^{2}$ by resolving the
following recursion relations for $u$, $v$ and $w$ written to that order

\begin{eqnarray}
u^{\prime } &=&b^{\epsilon -2\eta }\left\{ u-K_{d}\ln {b}\left( 1+\frac{%
\epsilon }{2}\ln {b}\right) \left[ \frac{n+8}{2}u^{2}-6u(v+w)\right]
-3uw(\epsilon -\delta )K_{4}\ln ^{2}{b}\right.  \nonumber \\
&&+K_{4}^{2}\ln ^{2}{b}\left[ \frac{1}{4}(n^{2}+6n+20)u^{3}-\frac{3}{2}%
(n+8)u^{2}(v+w)+9u(v+w)^{2}\right]  \nonumber \\
&&+\left. K_{4}^{2}\ln {b}(1+\ln {b})\left[ \frac{1}{2}%
(5n+22)u^{3}-6(n+5)u^{2}(v+w)+21u(v+w)^{2}\right] \right\} ,  \label{eqB}
\end{eqnarray}

\begin{eqnarray}
v^{\prime } &=&b^{\epsilon -2\eta }\left\{ v-K_{d}\ln {b}\left( 1+\frac{%
\epsilon }{2}\ln {b}\right) \left[ (n+2)uv-2v(v+w)-2(v+w)^{2}\right]
-3vw(\epsilon -\delta )K_{4}\ln ^{2}{b}%
\vphantom{\frac{3}{4}}%
\right.  \nonumber \\
&&-2w^{2}(\epsilon -\delta )K_{4}\ln ^{2}{b+}K_{4}^{2}\ln ^{2}{b}\left[
2(v+w)^{3}+3v(v+w)^{2}+\frac{3}{4}(n+2)^{2}u^{2}v-3(n+2)uv(v+w)\right] 
\nonumber \\
&&+K{_{4}}^{2}\ln {b}(1+\ln {b})\left[ 8(v+w)^{3}+3v(v+w)^{2}+\frac{3}{2}%
(n+2)u^{2}v-3(n+2)u(v+w)^{2}\right.  \nonumber \\
&&-\left. \left. 
\vphantom{\frac{3}{4}}%
3(n+2)uv(v+w)\right] \right\} ,  \label{eqC}
\end{eqnarray}
\begin{eqnarray}
w^{\prime } &=&b^{\delta -2\eta }\left\{ w-K_{d}\ln {b}\left( 1+\frac{%
\epsilon }{2}\ln {b}\right) [(n+2)uw-2w(v+w)]-w^{2}(\epsilon -\delta
)K_{4}\ln ^{2}{b}%
\vphantom{\frac{3}{4}}%
\right.  \nonumber \\
&&+K_{4}^{2}\ln ^{2}{b}\left[ 3w(v+w)^{2}+\frac{3}{4}%
(n+2)^{2}u^{2}w-3(n+2)uw(v+w)\right]  \nonumber \\
&&+\left. K_{4}^{2}\ln {b}(1+\ln {b})\left[ 3w(v+w)^{2}+\frac{3}{4}%
(n+2)u^{2}w-3(n+2)uw(v+w)\right] \right\} .  \label{eqD}
\end{eqnarray}

In (\ref{eqB}--\ref{eqD}), $b\geq 1$ is a rescaling factor, and all
integrals over momenta were taken within the shell $b^{-1}<k<1$. The
constant $K_{d}=2^{-(d-1)}\pi ^{-d/2}/\Gamma (\frac{d}{2})$ will be later
absorbed in the definition of the f.p. values $u^{*}$, $v^{*}$ and $w^{*}$,
in correspondence with (\ref{eq11}--\ref{eq13}). Notice also that for
deriving the above recursion relations we ignored the mass-renormalization
graphs \cite{Lub,Aharony,Bruce}. Putting $w=0$ in (\ref{eqB}--\ref{eqD}) one
immediately obtains the recursion relations for the SR case \cite{Lub,GMM}.

After some algebra, including correctly all terms up to order $\epsilon ^{3}$%
, one can see that all the $\ln {b}$ contributions in the recursion
relations cancel when one calculates the f.p. values of the parameters,
leading to the following expression for $v^{*}+w^{*}$ (here we write only
its value because of the large and complicated f.p. expressions for the
other parameters)

\begin{eqnarray}
v^{*}+w^{*} &=&\frac{1}{2(5n+4)}\left( (n+8)\delta -2(n+2)\epsilon \right) +%
\frac{1}{16(5n+4)^{3}}\left[ 9(15n^{3}+76n^{2}+16n+64)\delta ^{2}\right. 
\nonumber \\
&&+\left. 2(-95n^{3}-108n^{2}+240n-64)\epsilon \delta
-32(n+2)^{2}(5n+2)\epsilon ^{2}\right] .  \label{eq23}
\end{eqnarray}

Substituting this last expression in (\ref{eq16}), we finally obtain up to
order $O(\epsilon ^{2},\epsilon \delta ,\delta ^{2})$%
\begin{eqnarray}
z &=&2+\frac{1}{2(5n+4)}\left( (n+8)\delta -2(n+2)\epsilon \right) +\frac{%
(n+2)}{4(5n+4)^{3}}\left[ 3(n-4)(5n-8)\delta (\delta -\frac{2}{3}\epsilon
)\right.  \nonumber \\
&&-\left. 
\vphantom{\frac{2}{3}}%
4(10n^{2}+19n+4)\epsilon ^{2}\right] +\frac{3(n+2)}{(5n+4)^{2}}\ln \left( 
\frac{4}{3}\right) \left[ \epsilon ^{2}-6\epsilon \delta +9\delta
^{2}\right] .  \label{eq24}
\end{eqnarray}
which is valid for $n>1$. The case $n=1$ should be treated separately
because of the degeneracy of the recursion relations \cite{GL,GMM,WH}.
According to \cite{WH}, the inclusion of the term $w\neq 0$ breaks the
degeneracy, and in order $O(\epsilon )$ the LR disorder f.p. found before is
a solution of the recursion relation for $n=1$. It is however unstable for $%
\delta \sim O(\epsilon )$. Performing thus an expansion in $\epsilon ^{1/2}$%
, and assuming $\delta =O(\epsilon ^{1/2})$, one obtains for the f.p. values 
\cite{WH} 
\begin{equation}
u^{*}=\frac{2}{3}\delta \;;\;v^{*}=\frac{\delta }{2}\;;\;w^{*}=O(\epsilon ),
\label{eq25}
\end{equation}
which leads to results consistent with the extended Harris criterion.

With the aid of (\ref{eq25}), we get from (\ref{eq16}) the dynamic exponent
for the case $n=1$, 
\begin{equation}
z=2+\frac{\delta }{2}+O(\epsilon ).  \label{eq26}
\end{equation}

Equations (\ref{eq24}) and (\ref{eq26}) complete our study of the critical
dynamics of the TDLG equations in the presence of LR correlated quenched
impurities. A further investigation could be the study of the regions of
stability of the LR critical behavior taking into account the $\epsilon ^{2}$%
corrections. This, however, will not affect the qualitative behavior of the
phase transition, and a similar picture to that described in \cite{WH},
characterized by oscillatory corrections to the scaling, will appear.

In summary, in this work we investigated the dynamic scaling of the TDLG
equations with non-conserved order parameter (Model A) in the presence of
quenched impurities with LR spatial correlations. Using a double expansion
in two small parameters $\epsilon $ and $\delta \sim O(\epsilon )$, we
calculated the critical exponent $z$ up to order $O(\epsilon ^{2})$. We
showed that the quenched impurities of this kind affect the critical
dynamics already in first order in $\epsilon $ and $\delta $, which is a
common characteristic for systems with quenched impurities, independently of
the kind of correlations (short- or long-range).

This work is supported by DGICYT project PB94-0388, NATO grant CRG95055 and
partially by Contract F-608 from the Bulgarian Scientific Foundation.

\end{document}